\documentclass[journal,comsoc]{IEEEtran}

\usepackage{graphicx} 
\usepackage{epstopdf}
\usepackage[caption=false,font=footnotesize]{subfig}
\usepackage{color}
\usepackage{multirow}
\usepackage{array}
\usepackage{cite}
\usepackage{amsmath}
\usepackage{amsfonts}
\usepackage{amssymb}

\usepackage{algorithm}
\usepackage[noend]{algpseudocode}
\usepackage{nomencl}

\makenomenclature

\begin{document}
\title{Delay and Communication Tradeoffs for Blockchain Systems with Lightweight IoT Clients}

\author{Pietro Danzi,~\IEEEmembership{Student Member,~IEEE,} Anders E. Kal{\o}r,~\IEEEmembership{Student Member,~IEEE,} \\ \v{C}edomir Stefanovi\'c,~\IEEEmembership{Senior Member,~IEEE,} Petar Popovski,~\IEEEmembership{Fellow,~IEEE}
\thanks{Authors are with the Department of Electronic Systems, Aalborg University, Denmark, Email: \{pid, aek, cs, petarp\}@es.aau.dk.}
\thanks{The work was supported in part by the European Research Council (ERC Consolidator Grant no. 648382 WILLOW) within the Horizon 2020 Program.}
}

\markboth{Submitted for publication (2018)}
{Danzi \MakeLowercase{\textit{et al.}}: Delay and Communication Tradeoffs in Blockchains with Aggregated Updates for Lightweight Clients}
\maketitle

\maketitle

\begin{abstract}

The emerging blockchain protocols provide a decentralized architecture that is suitable of supporting Internet of Things (IoT) interactions.
However, keeping a local copy of the blockchain ledger is infeasible for low-power and memory-constrained devices. For this reason, they are equipped with \emph{lightweight} software implementations that only download the useful data structures, e.g. state of accounts, from the blockchain network, when they are updated.
In this paper, we consider and analyze a novel scheme, implemented by the nodes of the blockchain network, which aggregates the blockchain data in periodic updates and further reduces the communication cost of the connected IoT devices.
We show that the aggregation period should be selected based on the channel quality, the offered rate, and the statistics of updates of the useful data structures.
The results, obtained for the Ethereum protocol, illustrate the benefits of the aggregation scheme in terms of a reduced duty cycle of the device, particularly for low signal-to-noise ratios, and the overall reduction of the amount of information transmitted in downlink (e.g., from the wireless base station to the IoT device).
A potential application of the proposed scheme is to let the IoT device request more information than actually needed, hence increasing its privacy, while keeping the communication cost constant.
In conclusion, our work is the first to provide rigorous guidelines for the design of lightweight blockchain protocols with wireless connectivity. 
\end{abstract}


\section{Introduction}

\IEEEPARstart{S}{ince} the advent of the Bitcoin protocol in 2008~\cite{nakamoto2008bitcoin}, a large wave of blockchain protocols has emerged, aiming to support the implementation of decentralized applications, or \emph{dApps}, that reduce the need of a central authority to supervise the interactions in multi-agent systems~\cite{christidis2016blockchains}.
A promising use of dApps is in the Internet of Things (IoT) services, e.g. for executing economic transactions in smart grid applications in which devices interact through a blockchain to trade energy~\cite{danzi2017distributed,horta2017novel} or to solve distributed optimizations~\cite{munsing2017blockchains}.

\begin{figure}[!tb]
\centering
 {\includegraphics[width=0.7\columnwidth]{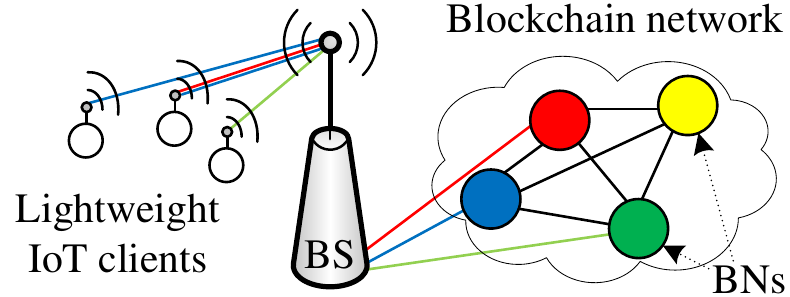}}
  \caption{Communication architecture for the interaction of IoT devices with a set of blockchain nodes (BNs) via wireless links provided by a base station (BS).}\label{fig:arch}
\end{figure}

Storing the entire blockchain and processing every transaction require a remarkable amount of storage memory and computations.
This is not feasible for IoT devices, as they are often constrained with respect to memory, computation, communication and power.
Instead, the IoT devices may act as \emph{lightweight} clients, which only store a subset of the blockchain data, possibly encoded~\cite{perard2018erasure}, needed to verify certain events of interest.
The IoT devices, configured as lightweight clients, are connected to a set of regular blockchain nodes (BNs), e.g. via a wireless base station, see Fig.~\ref{fig:arch}.
These devices are constantly synchronizing with the BNs~\cite{danzi2017analysis}, receiving a minimal amount of information, namely the \emph{block headers}.
In addition, when certain events that are of interest to a specific device occur, e.g. modification of specific accounts' state or transactions involving these accounts, the BNs transmit the updates to the device, including the proof of their inclusion (PoIs) in the blockchain.

While the architecture with lightweight clients reduces the processing and memory requirements, there is still a need for a remarkable amount of downlink traffic in order to maintain synchronization to the global blockchain~\cite{danzi2017analysis}. 
This type of operation challenges the common assumption that IoT devices mostly generate uplink traffic \cite{shafiq2012first, bockelmann2016massive}, urging the investigation in accurate models for blockchain traffic.
Schemes that reduce the amount of traffic exchanged between the lightweight clients and the BNs have previously been proposed, either by modifying the block structure \cite{kiayias2017non}, \cite{palai2018empowering}, by leveraging on the characteristics of account-based blockchains, e.g. Ethereum \cite{devto}, or by backing the authenticity of the transmitted information with a deposit of credit \cite{slockit}, \cite{gruberunifying}.

This work is motivated by the observation that the blockchain synchronization process can be tailored to the actual requirements of timely information updates to the IoT devices.
That is, the ultimate target is not to keep the devices always synchronized, but to synchronize them according to the needs of the underlying dApp.
Hence, we replace the legacy scheme, in which the BNs transmits the information to the devices whenever available, with a novel approach in which the information is accumulated and pushed only when needed by the end IoT devices.
Among the multitude of blockchain protocols, we focus on the Ethereum specification~\cite{wood2014ethereum}, but the overall principle of aggregation can be applied to other ``account-based" blockchains.

\begin{figure*}
\centering
  \includegraphics[width=2\columnwidth]{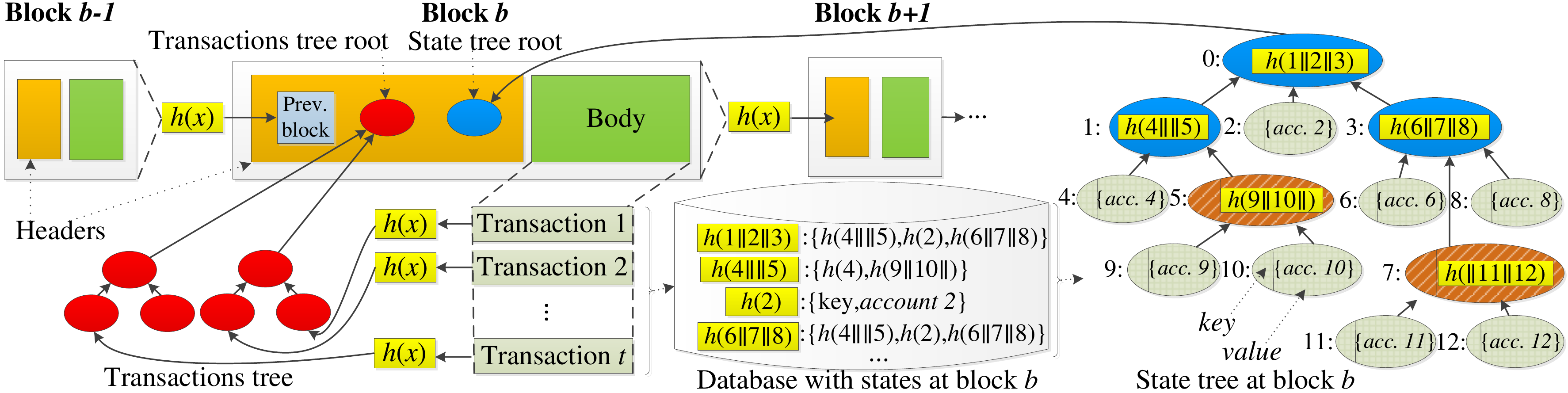}
\caption{Example structure of a blockchain. $h(x)$ is the hash value of node $x$ and $||$ is the concatenation operation. The $t$ transactions included in block $b$ apply modifications to the accounts' states, which are stored in a database. The state tree depicted on the righthand side, ternary in this example, is build from this database, and its root included in the block header. Branch nodes are colored in blue and extension nodes in brown. Leaf nodes (there are eight of them in the example) are composed by key and value, and colored in green.
}\label{fig:block}
\end{figure*}

The contributions of this work can be summarized as follows:
\begin{enumerate}
\item We propose and analyze an aggregation scheme, implemented at the BNs, that reduces the duty cycle of the device and the amount of transferred data, at the cost of an increased information delay at the IoT device. The reduction is achieved when events of interests to the device occur multiple times within an aggregation period, and is mainly caused by avoiding transmission of temporary states, but also because the size of the proof of inclusion increases sublinearly with the number of events.
\item We extend our previous model~\cite{danzi2017analysis} by including the possibility for the IoT device to observe multiple accounts, and for the base station to select the transmission rate. 
The result is a model for lightweight clients that is rich, but simple to analyze. We show its potential application by constructing a set of observed accounts that increases the privacy of the IoT device, while keeping the communication cost low.
\item We study the cost of transmitting the proof of inclusion for the updated data, namely the Merkle-Patricia tree data structures, and provide experimental results obtained for Ethereum protocol.
\end{enumerate}

The remainder of the paper is organized as follows.
Section~\ref{sec:protocol} provides an introduction to blockchain protocols, focusing on Ethereum, and describes the lightweight protocol variants.
The system model is introduced in Section~\ref{sec:model} and analyzed in Section~\ref{sec:analysis}.
Section~\ref{sec:results} presents the evaluation and Section~\ref{sec:conclusion} concludes the paper.

\section{Blockchain protocol}\label{sec:protocol}

This section introduces the main components of the Ethereum protocol \cite{wood2014ethereum}, that is a popular choice for blockchain systems tailored for IoT applications \cite{danzi2017distributed,8386853,alphand2018iotchain}. The common trait of Ethereum with other blockchain protocols can be found in~\cite{narayanan2016bitcoin}.

The Ethereum blockchain is a database that records of history of the states of \emph{accounts} in a chain of blocks.
An account is a data structure that contains an amount of credit and a general-purpose memory block.
The account may also contain a set of predefined procedures that can read and write to the memory; in this case, the account is called \emph{smart contract}.
The state of an account can be changed by \emph{transactions}, either directly or through the invocation of a procedure in a smart contract.
We shall refer to these modifications of accounts as \emph{events}.

Transactions are signed by devices using an asymmetric cipher, and identified by their hash values\footnote{The hash value of some input data $x$ is the output of a hash function defined by the blockchain protocol, and is indicated as $h(x)$.}, as in the Bitcoin specification~\cite{nakamoto2008bitcoin}.
The transactions are organized in a chain of blocks.
Besides a set of transactions, each block contains cryptographic signatures of the current states of the accounts and a pointer to the preceding block in the chain, which defines a causal relationship between blocks.
When a block is appended to the blockchain, the transactions that it includes are considered valid.

The Ethereum database is replicated at multiple nodes that are interconnected by a communication network.
Each node can append new blocks to the blockchain and inform the rest of the network about a new state of the database.
To avoid uncontrolled generation of blocks, so as to keep the database replications consistent, the nodes establish a leader election mechanism that (i) reduces the probability that more than one node generates a valid block at the same time and (ii) keeps the block generation rate lower than the propagation delay of the communication network.
Notable examples of leader election mechanisms in blockchains are Proof of Work~\cite{nakamoto2008bitcoin} and Proof of Stake~\cite{kiayias2017ouroboros}.
In order for a block to be considered valid, it must provide information that can be used to verify that it has been generated in accordance with the leader election mechanism.

\subsection{The block data structure}

A block is composed of a header and a body, see Fig.~\ref{fig:block}.
The block header has a fixed size, while the rest of the block contains the actual transactions and has a variable size.
When the number of transactions in a block is high, the variable-size part takes a dominant portion of the total block size.
The information specified in the header includes: the block hash value, an incremental counter, the cryptographic signature of the node that generated it, the proof that the block is valid, e.g., Proof of Work solution, and one or more hash values that represent roots of PoI trees. In this work, we mainly consider the transactions tree and the state tree.
The transactions tree, which uniquely binds the modifications of accounts certified by a block with the block header, can be used to prove that specific transactions are included in the block.
The state tree, depicted on the right-hand side of Fig.~\ref{fig:block}, provides a snapshot of the entire collection of account states.\footnote{The state tree is a characteristic of the Ethereum specification.}

\subsection{Proof of inclusion (PoI) via Merkle-Patricia trees}\label{sec:poi}
\begin{figure}[!tb]
\centering
  \subfloat[]{\includegraphics[width=0.24\columnwidth]{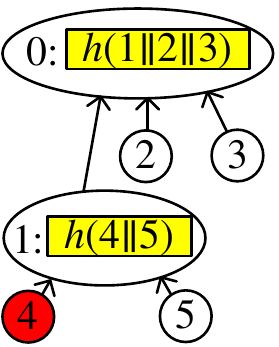}}
  \hfill
  \subfloat[]{\includegraphics[width=0.26\columnwidth]{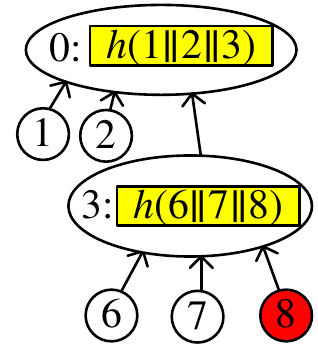}}
    \hfill
  \subfloat[]{\includegraphics[width=0.35\columnwidth]{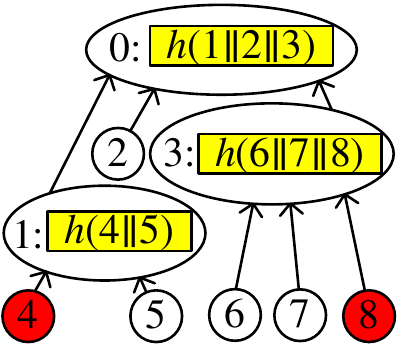}}
    \caption{Representation of the PoIs (a) of node 4, (b) of node 8 of Fig.~\ref{fig:block}, and (c) PoMI of 4 and 8. $h(x)$ is the hash value of $x$ and $||$ is the concatenation operation.}
    \label{fig:pois}
\end{figure}

The Ethereum protocol provides PoIs using Merkle-Patricia trees~\cite{wood2014ethereum,merkling,albassam2018fraud}.
A Merkle-Patricia tree has three types of nodes; leaf, extension and branch nodes, see Fig.~\ref{fig:block}, as in standard Patricia trees \cite{morrison1968patricia}, and is used to efficiently store and retrieve data structures associated with strings.
In the blockchain context, the string is the hash value of the address of an account or transaction, and the data structure to be retrieved is the account/transaction itself.
The branch nodes only store the hash value of the list of its child nodes, see Fig.~\ref{fig:block}.
Leaf and extension nodes, also illustrated in Fig.~\ref{fig:block}, store a key, that is the hash value of the common path shared by all child nodes, and a value.
The value stored by extension nodes is the hash value of the list of child nodes, and the one of the leaf nodes is the hash value of the data that is to be authenticated (e.g. an account or transaction).
The use of a hash function to index the addresses provides equal length of the strings, which are equiprobable.

The presence of a specific node in the Merkle-Patricia tree is proven by constructing its PoI.
A PoI is a collection of node values that enables generation of the hash value, contained by the root node of the tree, e.g. the node labeled $0$ in Fig.~\ref{fig:block}, starting from the specific node to prove.
By comparing the generated root hash value with the value stored in the block header, the inclusion in the blockchain of the data structure, associated with the specific node, can be verified \cite{nakamoto2008bitcoin}.
In practice, the PoI is used to verify that a particular leaf node, i.e. an account or transaction, is present in a state tree.
Specifically, a PoI is created by starting from the root of the Merkle-Patricia tree, and descending to the specific node. 
At each level, all nodes, that are siblings to the node on the path from the root to the specific node, are collected, as illustrated in Fig.~\ref{fig:pois} ((a) and (b)).
Notice that a PoI, in general, contains much fewer nodes than the complete Merkle-Patricia tree since most of the branches are not collected during the descent from the root \cite{albassam2018fraud}.

A single proof can be constructed to prove multiple data structures by collecting the union of the nodes required to prove each of the data structures.
We shall refer to such a proof as a Proof of Multiple Inclusions (PoMI)\footnote{In contrast with prior literature \cite{merkling}, we use the terms PoI and PoMI to differentiate the proof from the blockchain-specific data structure that provides it, e.g. Merkle tree (in Bitcoin) or Merkle-Patricia tree (in Ethereum).}.
Since the nodes required to prove each data structure are likely to intersect, a PoMI is typically much smaller than if each data structure is to be proven by an individual PoI. Fig.~\ref{fig:pois}(c) provides an example of this reduction, for the proof of both nodes 4 and 8. If two individual proofs are build, the PoI of node 4 contains nodes \{5, 2, 3\} and the PoI of node 8 contains \{1, 2, 6, 7\}, such that seven nodes are needed in total. However, if the proofs are sent together in a PoMI, only nodes \{2, 5, 6, 7\} are needed, motivating the advantage of using this data structure.

\subsection{Synchronization protocols}\label{sec:syncprotocols}
\begin{figure}[!tb]
\centering
  \subfloat[]{\includegraphics[width=\columnwidth]{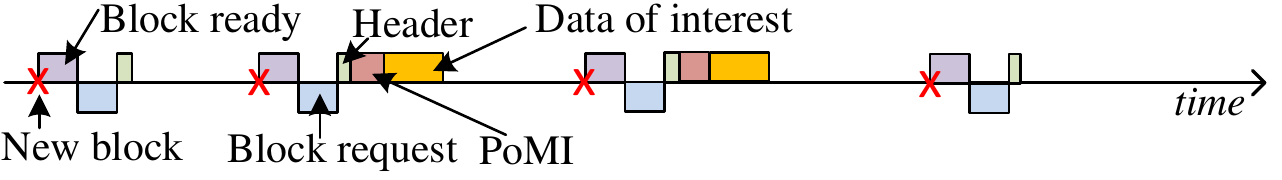}}
  
  \subfloat[]{\includegraphics[width=\columnwidth]{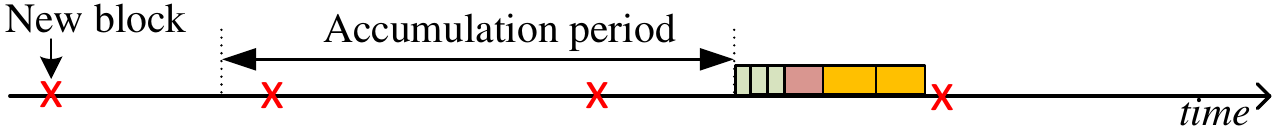}}
    \caption{Information exchanged between a client and a BN using a lightweight protocol during four block periods, (a) without aggregation scheme and (b) with aggregation. Downlink/uplink messages are depicted above/below the time arrow. ``Data of interest" includes the accounts' data and relative PoMIs.}
    \label{fig:time}
\end{figure}

A blockchain client is updated on modifications of the blockchain database, observed by BNs, by means of a synchronization protocol.
In~\cite{danzi2017analysis} we have presented three possible protocols that can be adopted for this purpose, denoted by P1, P2 and P3.
With P1, the client itself stores the entire blockchain, and locally checks the correctness of the transactions.
This configuration is not envisaged for IoT devices due to the requirements of storage memory and processing, and will not be considered in this work.
In P2, the BNs are notified about the account updates that the client is interested in receiving. Hence, the client receives the block headers from the BNs, by default, and the accounts of interest, only when they are modified.
This scheme, referred to as a \emph{lightweight protocol}, reduces the amount of data communicated in the downlink, as well as the amount of local processing.
In fact, the client only verifies that the information sent by BNs is consistent, delegating to them the auditing of the actual validity of the transactions \cite{albassam2018fraud}. It follows that, with P2, the client must be connected to at least one honest BN to be able to detect the presence of false information.
Finally, in protocol P3 the IoT device is connected to a proxy node that only sends to the device the useful information, without providing any proof that is included in the blockchain.
This protocol is also not considered in this work, as it requires the device to fully trust the proxy node, which is not in line with the envisioned trustless decentralized architecture.

In this paper, we assume that the IoT devices have memory and processing limitations and only consider the lightweight protocols of type P2, such as Bitcoin's Simplified Payment Verification (SPV)~\cite{nakamoto2008bitcoin} and the Ethereum Light Client~\cite{zsfelfoldi}, which are designed to allow clients to verify the inclusion of transactions in the blockchain without downloading and storing the entire blockchain.
Specifically, clients only download block headers, as well as the data structures (e.g. account or transaction data structures) that are of interest.

Fig.~\ref{fig:time} shows the messages exchanged between a client and a BN using a lightweight protocol during three block periods. 
The red crosses represent the instants at which new blocks are generated.
In the basic lightweight protocol, Fig.~\ref{fig:time}(a), the information is pushed in the downlink from BS as it becomes available.
In the first block period there is no information of interest, and only the block header is sent, while in the second and third periods there are events of interest and the respective data are sent with their PoMI.

The presence of both transactions and state tree roots in the Ethereum block headers, see Fig.~\ref{fig:block}, permits to adopt two different approaches to update the local copy of the account states, as illustrated with the following example.
Suppose that an account is updated multiple times during several block periods. 
The BN can send the last version of the account data, with the corresponding partition of \emph{state} tree at the last block.
In this case, the IoT device just replaces the local data if the PoI root matches the one included in the last block header, otherwise refuses it, cf.~\cite{devto, merkling}.
Alternatively, the BNs send the whole sequence of transactions that modified the account, along the block periods, together with the collection of their PoIs build from the \emph{transactions} tree.
The sequence of transactions is applied, by the device, to its local version of the account state, to finally reconstruct the updated state.
In this paper, we only consider the first approach, and we remark the extension to the second one in Sec.~\ref{sec:remark}.

\section{System Model}\label{sec:model}

The scheme proposed in this paper aggregates the information in order to reduce the communication cost, see Fig.~\ref{fig:time}(b).
Given that the application run by the client can tolerate a delay, information is accumulated at the BN and then periodically released at the subsequent aggregation point.
The approach followed by the BN is to always send a proof by means of the state tree, triggering the replacement of the local copy of the client.
This permits to send only the \emph{latest} version of accounts that are modified multiple times during the accumulation, and merge the PoIs of accounts modified in \emph{different} blocks, in a unique PoMI.
The scheme is investigated in detail in the rest of the paper.

\subsection{Blockchain network and IoT device}

We consider a blockchain network in which new blocks are generated at exponentially distributed intervals with (network-wide) rate $\lambda$.
A single IoT device is connected to a set of $N$ BNs via a wireless link through a base station, see Fig.~\ref{fig:arch}.
The device subscribes to block headers for all generated blocks as well as state updates for a set $\mathcal{A}$ of accounts, which are a subset of the existing accounts.
The generic account, indexed as $j \in \mathbb{N}$, is updated independently in a block with probability (or relative frequency) $p_j$.
We consider the case where the device is not interested in the full state history of the observed accounts, but merely in their most recent state.
That is, the device needs to be informed about only the most recent state of the observed accounts, as well as receive the PoMI that proves the inclusion of the specific account states in the blockchain.
The case is representative of a class of problems in which the \emph{age} of the information, i.e. data freshness, is more valuable than tracking all state changes, and includes environmental monitoring applications and power grid stabilization systems~\cite{updateorwait}.

To simplify the presentation, we assume that a block header and updated accounts' states take up a fixed number of $l_{\text{H}}$ and $l_{\text{a}}$ bits, respectively. In contrast with this, the size of the PoMI, with length $l_{\text{PoMI}}$ bits, is random as a result of the PoMI tree data structure.
Specifically, as described in Section~\ref{sec:protocol}, the size of the PoMI is sublinear in the number of accounts.

\subsection{Aggregation protocol}

The block headers and the updated observed accounts are aggregated at a BN, termed \emph{aggregation BN}, selected by the device, and transmitted to the device periodically every $T$ seconds.
The value of $T$ depends on the information delay, tolerated by the application, from the instant at which the account is modified, to the instant at which the update is delivered to the device.
Upon successful reception of the transmission, the IoT device acknowledges the packet.
We assume that the device selects the sequence of aggregation BNs, over different aggregation period, as part of the initial network association procedure, e.g. by means of a seed sequence. Consequently, the execution of the protocol only requires downlink messages, because all the information, needed by BNs, is sent by the IoT device in the initialization phase.
When no transmission is ongoing, the device is assumed to be in power-saving mode.

\subsection{Wireless link}

The wireless downlink from the base station to the IoT device is assumed to be a block Rayleigh-fading channel with constant channel gain over the duration of a transmission and independent channel gains across transmissions. 
This occurs, for example, in system based on per-packet frequency hopping (FH).
Due to the power constraints of the IoT device, we assume that the base station has no information about the channel and hence performs no power or rate adaptation.
As a result, a transmission may fail with probability~\cite{tse2005fundamentals}
\begin{equation}\label{eq:outage}
p_{\text{out}}= 1 - \exp{\left(-\frac{2^\frac{R}{W}-1}{\gamma}\right)},
\end{equation}
where $\gamma$ is the average received signal-to-noise ratio (SNR), $R$ is the transmission rate in bits/s, $W$ is the bandwidth of the channel in Hz.
The downlink packet is retransmitted until it has been received successfully by the device.

In contrast to the downlink transmissions, we assume that the transmission of the acknowledgment packet in the uplink happens instantaneously and is always received reliably, thanks to power control, performed at the IoT device side, based on the received transmission.

\subsection{Frame structure}

\begin{figure}[!tb]
\centering
 \includegraphics[width=0.7\columnwidth]{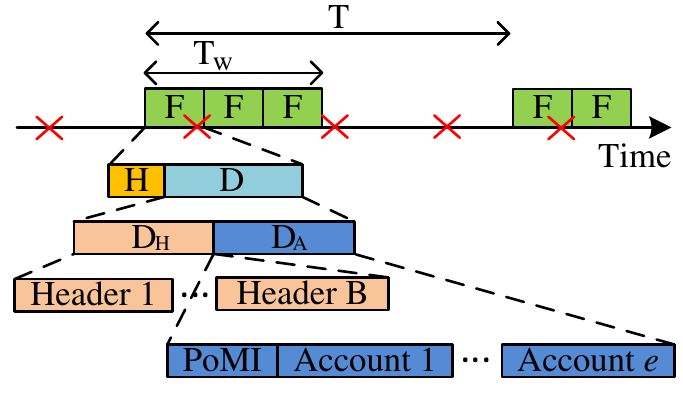}
  \caption{The periodic release of information. Red crosses represent block generations. In the first release, there are two retransmissions of the frame (F), due to failure, in the second release, only one retransmission.}\label{fig:periodic_release}
\end{figure}

The downlink frame, represented in Fig.~\ref{fig:periodic_release}, consists of $F$ bits, and is divided into a fixed number $H$ of header bits, representing the standard communication protocol overhead, and a variable number $D$ of payload bits, corresponding the blockchain information, i.e. $F=H+D$.
Its duration is related to the transmission rate $R$ as
\begin{equation}\nonumber
T_w = \frac{kF}{R}\qquad [\text{s}],
\end{equation}
where $k \geq 1$ is the number of transmissions, including retransmissions due to outage.

If the transmission of the frame takes longer than the transmission period, i.e. $T_w > T$, due to retransmissions, it is halted and considered failed.
In this case, that has been analyzed in \cite{danzi2017analysis}, since the block headers are required in order for the IoT device to stay synchronized to the blockchain, the next frame should include the block headers accumulated in the current frame.
In this work, we consider the channel and block generation parameters that provide a negligible probability that the frame cannot be received in the current transmission period, so that the phenomenon can be ignored.\footnote{Scenarios for which this assumption does not hold can be observed when the probability of outage is rather high, and the transmission rate is low.}
The $D$ payload bits are divided into $D_{\text{H}}$ bits for block headers and $D_{\text{A}}$ bits for account updates, i.e. $D=D_{\text{H}}+D_{\text{A}}$.
$D_{\text{H}}$ and $D_{\text{A}}$ are random as they depend on the number of generated blocks and account updates.

\section{Analysis}\label{sec:analysis}

In this section, we present an analysis of the aggregation scheme.
To this end, we first obtain the distribution of the frame size and frame transmission duration, and then use this result to evaluate the communication cost and latency.

\subsection{Frame size distribution}

Recall that the frame is divided into $H$ header bits, $D_{\text{H}}$ bits for block headers and $D_{\text{A}}$ bits for account states.
The $H$ header bits are fixed, while $D_{\text{H}}$ depends on the number of generated blocks during the aggregation period $T$, indicated as $B$.
Similarly, $D_{\text{A}}$ depends on the number of generated blocks, as it impacts the number of observed accounts that are updated.
We indicate the probability distributions of $D_{\text{H}}$ and $D_{\text{A}}$, conditioned on the number of generated blocks, respectively as $p_{D_\text{H}|B}$ and $p_{D_\text{A}|B}$.
As a result, we may factorize the distribution of the total frame size $F$ as
\begin{equation}\label{eq:f}
  p_{F}(f) = \sum_{b=0}^{\infty}p_{B}(b) \sum_{i=0}^f p_{D_\text{H}|B}(i|b)p_{D_\text{A}|B}(f-i|b),
\end{equation}
where we have used the fact that $f=D_{\text{H}}+D_{\text{A}}$. The possible sizes are conditioned on the event that $b$ blocks are generated during $T$ given by
\begin{equation}
  p_{B}(b)=\frac{(\lambda T)^b \exp({-\lambda T})}{b!}.
\end{equation}
The formula directly follows from the assumption of exponential waiting time between blocks, which has been shown to hold for blockchains based on Proof of Work~\cite{decker2013information}.

\subsubsection{Distribution of $D_{\text{H}}$}

Since we assumed that the block headers are always received within the current transmission period, i.e. $T_w < T$, the size of $D_{\text{H}}$, when $B$ blocks are generated, can be approximated with the fixed quantity $B\cdot l_{\text{H}}$ bits, yielding $p_{D_\text{H}|B}$ in \eqref{eq:f}.

If this is not the case, the number of block headers that needs to be transmitted should be modeled as a bulk queue, where blocks arrive according to a Poisson distribution with rate $\lambda$, and are
served in bulks of up to $\left\lceil \frac{D}{l_{\text{H}}}\right\rceil$, where $\lceil x\rceil$ denotes the smallest integer greater than or equal to $x$. However, this makes the accurate analysis intractable and is outside of the scope for this work. 

\subsubsection{Distribution of $D_{\text{A}}$}

In order to obtain the size of the account updates $D_{\text{A}}$, we first need to characterize the number of accounts $U$ that are updated during an aggregation period $T$.
The probability that account $j$, characterized by relative frequency $p_j$\footnote{This quantity is not assumed but experimentally estimated in Sec.~\ref{sec:results}.}, is updated at least once in $b$ blocks accumulated during the aggregation period is given by $q_j=1-(1-p_j)^b$.
We denote by $U$ the total number of accounts that are modified at least once in $B$ blocks.
Since each of the accounts is updated independently conditioned on $B$, $U$ follows a Poisson binomial distribution parameterized by the account update probabilities $q_1,q_2,\ldots,q_{|\mathcal{A}|}$:
\begin{equation}
p_{U|B}(u | b) = \sum_{\mathcal{B} \in \mathcal{F}_u} \prod_{j \in \mathcal{B}} q_j \prod_{l \in \mathcal{F}_u\setminus\mathcal{B}} \left( 1 - q_l \right).
\end{equation}
$\mathcal{F}_u$ is the set of all subsets of $\{1,2,\ldots,|\mathcal{A}|\}$ with cardinality $u$, $\mathcal{B}$ is an element of $\mathcal{F}_u$ and $\mathcal{F}_u\setminus\mathcal{B}$ is the complement of $\mathcal{B}$.
This distribution is used to find the distribution of $D_{\text{A}}$, conditioned on $B$:
\begin{equation}\label{eq:ppab}
p_{D_{\text{A}} | B }(a|b) = \sum_{u=0}^{|\mathcal{A}|} p_{D_{\text{A}}|U,B}(a | u , b) \cdot p_{U|B}(u | b).
\end{equation}

Notice that $p_{D_{\text{A}}|U,B}(a | u , b)$ only depends on the realization of the number of modified accounts $u$.
This permits us to write
\begin{align}
\nonumber p_{D_{\text{A}}|U,B}(a | u , b) &= p_{D_\text{A}|U}(a | u ) \\
\label{eq:p_a_distr} &= \sum_{i=0}^{a}p_{l_{\text{PoMI}}|U}(i | u) p_{l_{\text{acc}}|U}(a-i | u).
\end{align}
To complete the analysis, we need to characterize the size of PoMI and accounts' information.
The size of accounts' information varies only with the number $U$ of modified accounts,; as in our model they have fixed size of $l_{\text{a}}$, this size is simply $l_{\text{acc}} = U \cdot l_{\text{a}}$ bit.
On the other hand, the PoMI length $l_{\text{PoMI}}$ does not only depend on the number of observed accounts, but also on their position in the state tree.
For simplicity, we assume that the tree is perfectly balanced at all levels, and that the location of the observed accounts at the last level is uniformly distributed.
The approximation is supported by the fact that Patricia trees are generally well-balanced~\cite{rais1993limiting}.

However, the fact that the number of proofs that each node can be part of is bounded by the number of descendant leaf nodes, makes the problem of obtaining the distribution of the number of nodes in the PoMI a hard combinatorial problem; even the expected value of the number of nodes is computationally intractable to obtain.
Instead, we approximate the number of nodes by relaxing this condition. The resulting approximation captures the characteristics of the PoMI size as the number of modified accounts grows, and is accurate as long as the number of modified accounts is much lower than the total number of leaf nodes in the tree. This is typically the case, as the set of observed accounts is small.
Specifically, relaxation results in the following recursive approximation of the expected number of \emph{nodes} in a PoMI for $u$ accounts, when the tree has height $\eta$:
\begin{equation}\label{eq:n_eta}
\bar{N}_{\eta}(u) = \sum_{h=1}^{\eta} L\bar{N}_{h-1}(u)\left(1-\frac{1}{L\bar{N}_{h-1}(u)}\right)^u,
\end{equation}
with $\bar{N}_{0}(u) = 1$.
The derivation is given in Appendix~\ref{sec:appendix}.

To obtain the expected number of \emph{bits} for a PoMI, we assume that the tree does not contain extension nodes, as they have variable size, see~\cite{morrison1968patricia}.
Hence, with this approximation, the internal nodes are only branch nodes. Indicated the size of the output of the hash function with $l_{\text{s}}$, each internal node has fixed size of $l_{\text{s}}$ bits.
Instead, the leaf nodes are composed by a key and a value, see Sec.~\ref{sec:protocol}, both containing hash values, resulting in a fixed size of $2\cdot l_{\text{s}}$ bits.
In conclusion, we obtain the expected number of bits required for a PoMI of $u$ accounts:
\begin{equation}
\bar{l}_{\text{PoMI}}(u) = l_{\text{s}}\bar{N}_{\eta}(u) + u(2\cdot l_{\text{s}}).
\end{equation}

\subsection{Transmission duration}

The total transmission duration $T_w$ depends on $F$ and the number of (re)transmissions that is needed before the packet is successfully received by the IoT device. 

A frame is transmitted successfully with probability $1-p_{\text{out}}$, independent of the size of the frame, and hence the number of transmissions is geometrically distributed with the probability mass function
\begin{equation}
  \Pr(k\text{ transmissions}) = p_{\text{out}}^{k-1}(1-p_{\text{out}}).
\end{equation}
Since the rate remains fixed across (re)transmissions, it follows that the probability density function of $T_w$ is
\begin{align}\label{eq:p_t_w}
  p_{T_w}(t)&=\sum_{k=1}^{\infty}p_F\left(\frac{t}{k}R\right)\Pr(k\text{ transmissions})\\
  &=(1-p_{\text{out}})\sum_{k=1}^{\infty}p_F\left(\frac{t}{k}R\right)p_{\text{out}}^{k-1}.
\end{align}

\subsection{Data savings of the aggregation protocol}\label{sec:savings}

Since the states of accounts can be verified from the state tree root contained in the most recent block header, 
the aggregation scheme provides data savings by (i) sending a unique PoMI that certifies only the \emph{latest} state of the modified accounts, (ii) sending only the most updated copy of the account data structure, and (iii) reducing the amount of frame overhead, $H$.
We consider protocol P2 from~\cite{danzi2017analysis} introduced in Sec.~\ref{sec:syncprotocols} as the benchmark.
Recall that P2 requires the device to download the following information at each block period: the frame overhead $H$, a notification of the new block from each peer, the block header, the PoMI and the account data structures. 
In addition, with P2, the device receives a notification of new block, indicated as ``Block ready" in Fig.~\ref{fig:time}(a), from each BN, and consequently selects a BN with uplink message, indicated as ``Block request" in the same figure.
However, to establish a fair comparison with the aggregation protocol, we assume that the BN, in charge of sending the update, is pre-selected via random seed also in P2, removing the need of these messages.

The expected amount of bits downloaded with protocol P2 during a block period is
\begin{equation}\label{eq:cost_p2}
\mathbb{E} \left[ F^\text{(P2)} \right] =H + P = H + l_\text{H} + \sum_{a=0}^{\infty} a\cdot p_{D_\text{A}|B} (a| 1).
\end{equation}
The expression is based on \eqref{eq:ppab}, and on the fact that exactly one block is generated during a block period.

The expected number of bits per block period downloaded using the aggregation protocol proposed in this paper is given by averaging~\eqref{eq:f}:
\begin{equation}\label{eq:savings}
\mathbb{E}[F] = \frac{1}{\lambda\cdot T}\sum_{f=0}^{\infty} f\cdot p_{F} (f),
\end{equation}
where $\lambda\cdot T$ is the expected number of blocks within the aggregation period.
We can now express the savings of the aggregation protocol as
\begin{align}\label{eq:Gain}
\Gamma =& 1-\frac{\mathbb{E}[F]}{ \mathbb{E} \left[ F^\text{(P2)} \right]} =\\ 
=& 1-\frac{\sum_{f=0}^{\infty} f\cdot p_{F} (f)}{\lambda\cdot T\cdot\left(H + l_\text{H} + \sum_{a=0}^{\infty} a\cdot p_{D_\text{A}|B} (a| 1)\right)}.
\end{align}

\section{Evaluation}\label{sec:results}

\begin{table}[]
\centering
\caption{System parameters}
\label{table:params}
\begin{tabular}{llllll}
\hline
\multicolumn{6}{l}{Blockchain}                                                                                                                      \\ \hline
$\lambda$     & $0.1$ $\text{s}^{-1} $ & $H$    & 1200     		  & $l_\text{H}$ & 4046  \text{bit}    \\
$l_\text{a}$         & 320 \text{kb}          & $l_s$  & 256 \text{bit} & $L$   & 16                  \\
$\eta$ & $5$                  &   &  & & \\ \hline                           
\multicolumn{6}{l}{Communication channel}                                                                                                                      \\ \hline
$R$ & 250 kbit/s & $W$ & 180 kHz & $\gamma$ & 30 dB \\ \hline
\end{tabular}
\end{table}

To validate our model and show the performance of the aggregation scheme, we have modified the Python implementation of Ethereum protocol, PyEthereum \cite{pyethereum}.
The system, parametrized as listed in Table~\ref{table:params}, includes a randomly generated blockchain. This is obtained by generating accounts that contain random information, with size $l_\text{a}$ bits, and inserting them in a newly initialized blockchain database.
For the statistical characterization of accounts updates, we take as reference the Ethereum main network as described in the following section.

\subsection{Statistical characterization of accounts updates}
\begin{figure}[!tb]
\centering
\includegraphics[width=\columnwidth]{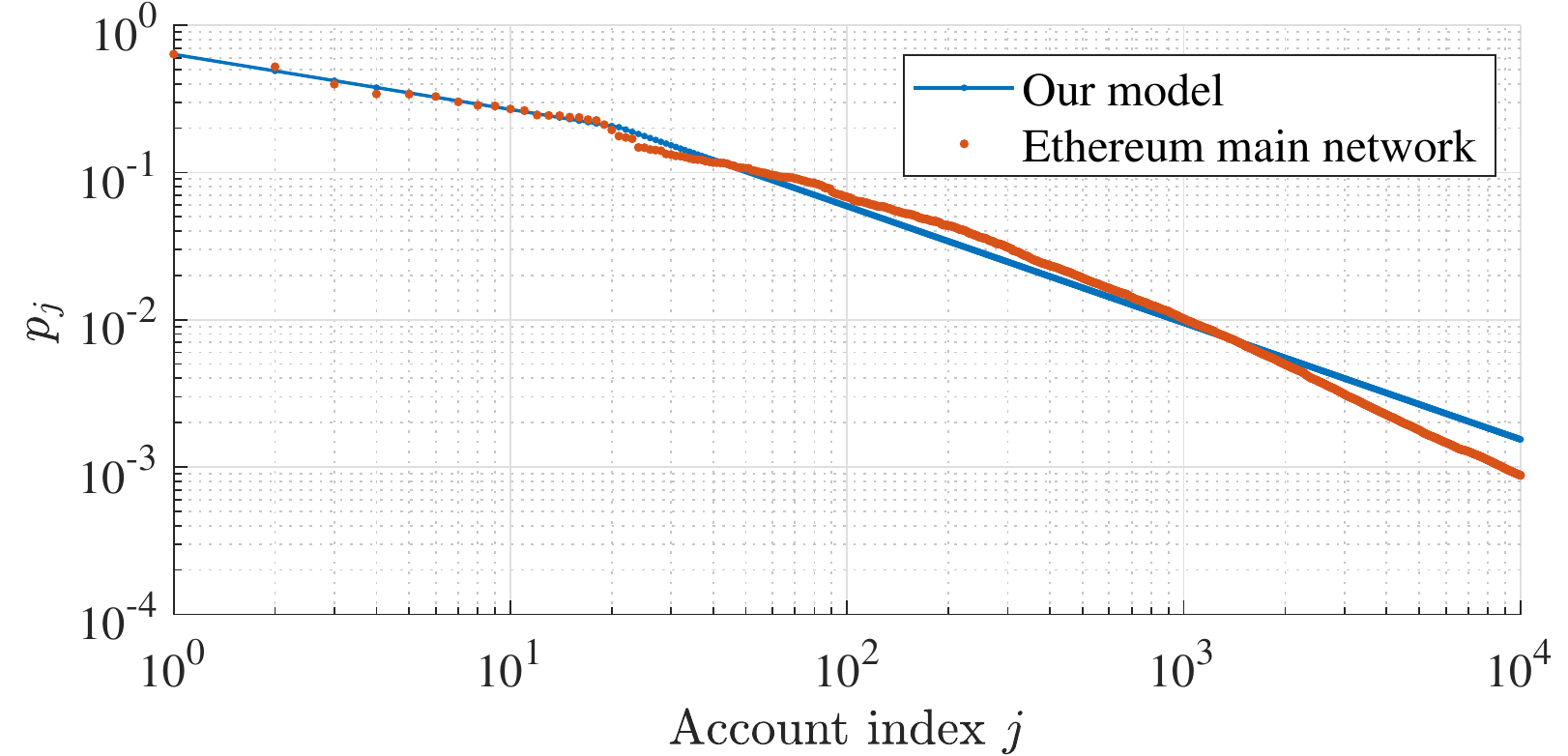}
  \caption{Relative frequency of updates for the most active accounts, in log-log scale.}\label{fig:fit}
\end{figure}

The statistics of account updates plays a fundamental role in the design and evaluation of blockchain protocols. We base our evaluation on the Ethereum main network dataset \cite{google}, by analyzing the activity during blocks numbered from 5.1 to 6.4 million.
Fig.~\ref{fig:fit} shows the frequency of updates of the $10^4$ most updated accounts, indexed in descending order of their updates frequencies. To extract this metric, we do not distinguish between transactions from/to the accounts, or consider if there are multiple transaction involving one account in the same block.
We model the relative frequency of updates of account $j$ according to the broken power-law:
\begin{equation}\nonumber
p_j =
\begin{cases}
\alpha_1 j^{\alpha_2} \quad &\text{if} \, j \leq \alpha_3,\\
\alpha_3^{\alpha_2 - \alpha_4} \alpha_1  j^{\alpha_4} &\text{otherwise}.
\end{cases}
\end{equation}
We opt for this function, instead of the plain power-law, adopted e.g. in \cite{lischke2016analyzing,somin2018social}, because the most frequent accounts are associated to web services that provide currency exchanges and are updated at similar rates.
The least squares fit gives $\alpha_1 = 0.63$, $\alpha_2 = -0.37$, $\alpha_3 = 21$, $\alpha_4 = -0.79$. This function, also shown in Fig.~\ref{fig:fit}, is used to generate the relative frequencies for our ``synthetic'' blockchain in the evaluation.

\begin{figure}[!tb]
\centering
\includegraphics[width=\columnwidth]{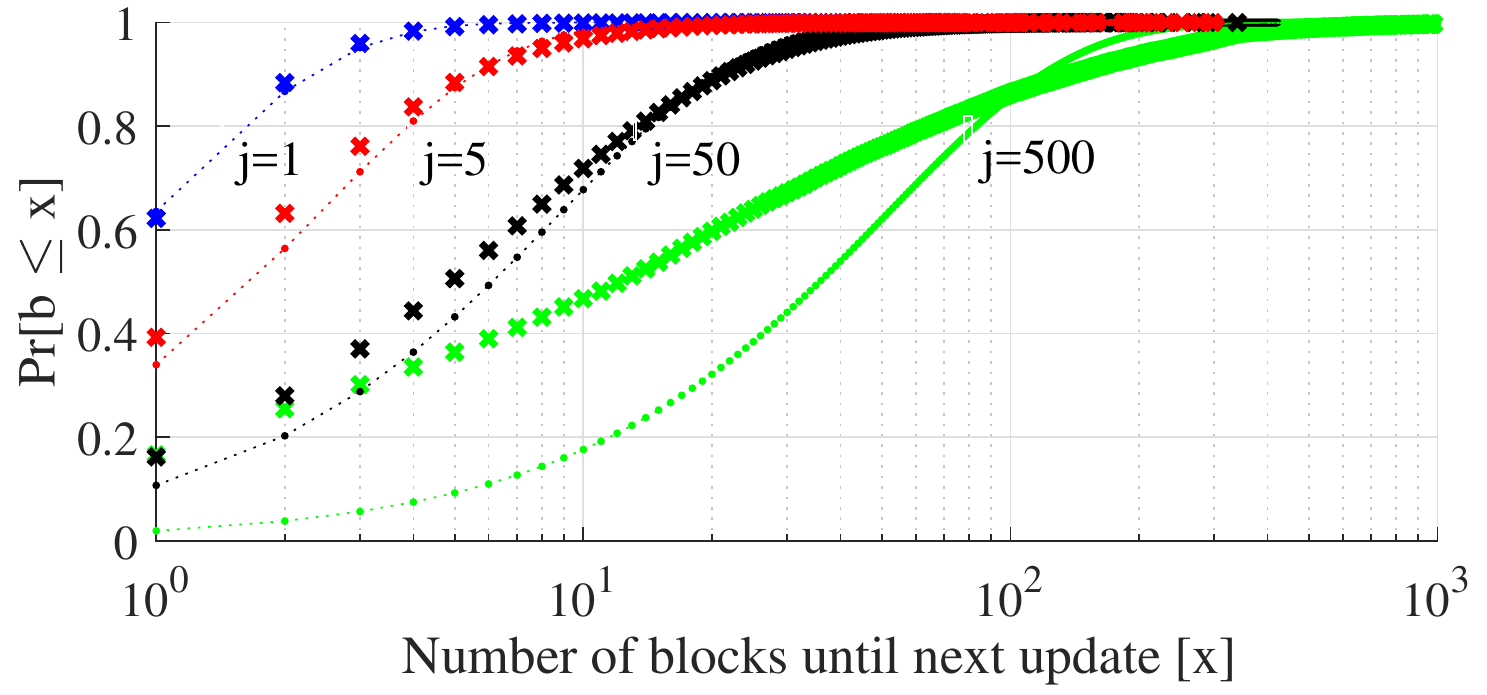}
  \caption{Comparison of empirical CDF of accounts (represented with crosses), with index $j$, with the CDF of geometrical distribution (represented with dots).}\label{fig:ecdf}
\end{figure}

In addition to obtaining the account update probabilities, we inspect the accuracy of modelling the number of blocks between two account updates as a geometric distribution as assumed in our model.
We compare the empirical cumulative density function (CDF) of an account, $j$, to the CDF of a geometric distribution with parameter $p_j$. Fig.~\ref{fig:ecdf} shows the results obtained for some accounts of the data set. It results that the assumption only holds for frequently updated accounts. This behaviour should be taken into account in future works.

\subsection{Validation of Merkle-Patricia proofs length}
\begin{figure}[!tb]
\centering
  \subfloat[]{\includegraphics[width=\columnwidth]{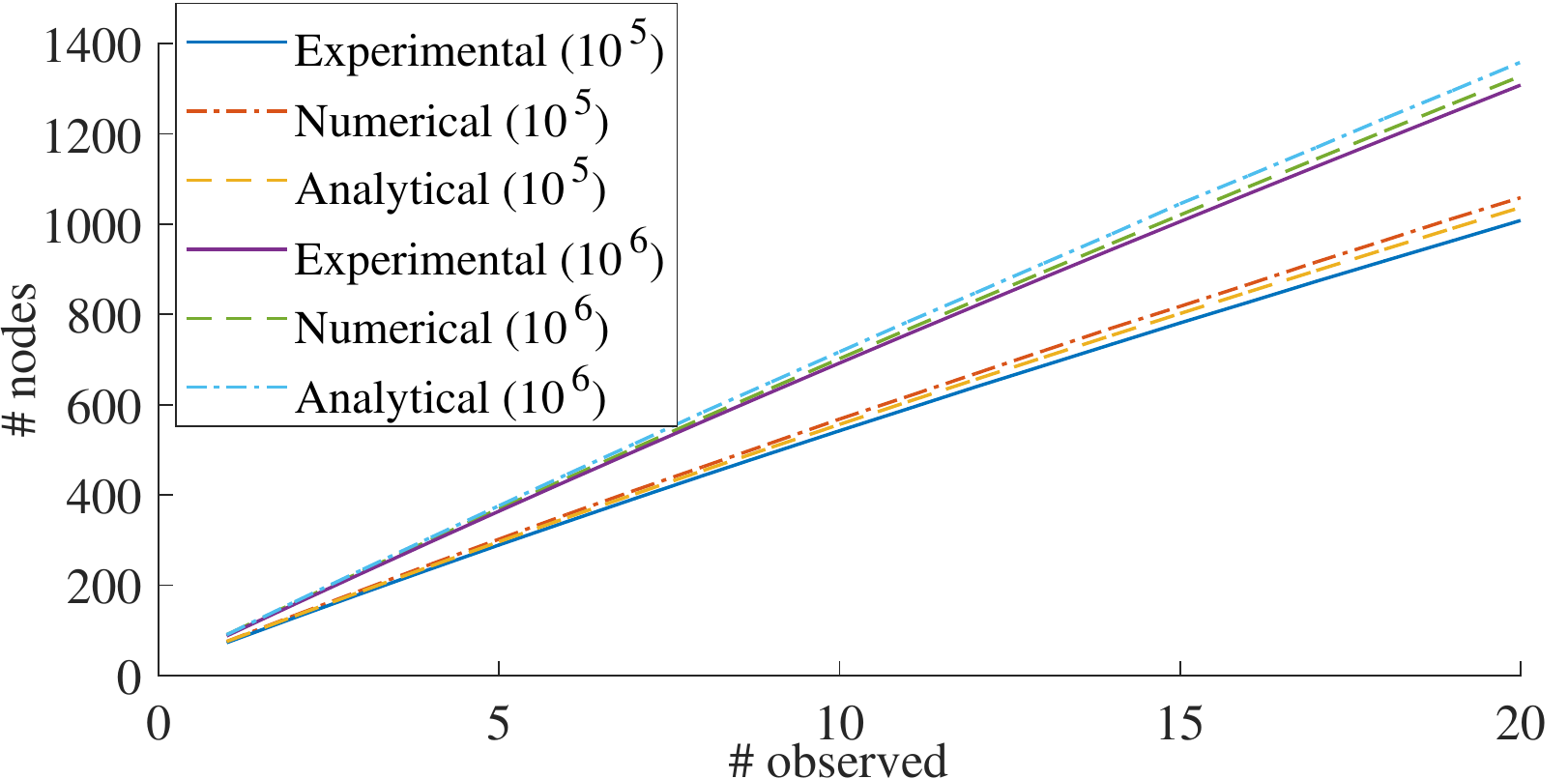}}
 
  \subfloat[]{\includegraphics[width=\columnwidth]{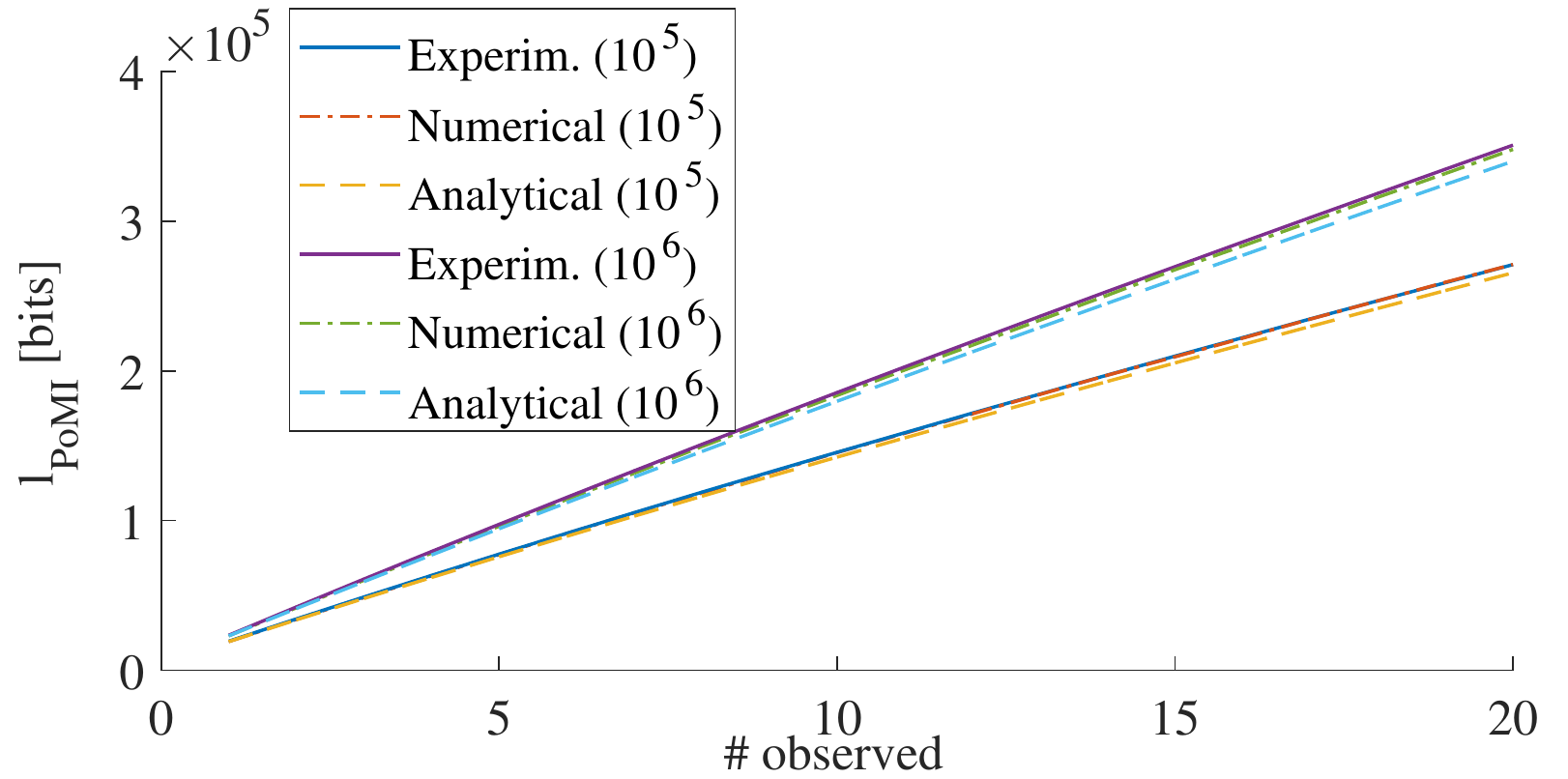}}
    \caption{Comparison of analytical approximation, numerical and experimental results for the
    (a) number of nodes needed in the PoMI and (b) its length.}
    \label{fig:ea}
\end{figure}
\begin{figure}[!tb]
\centering
\includegraphics[width=\columnwidth]{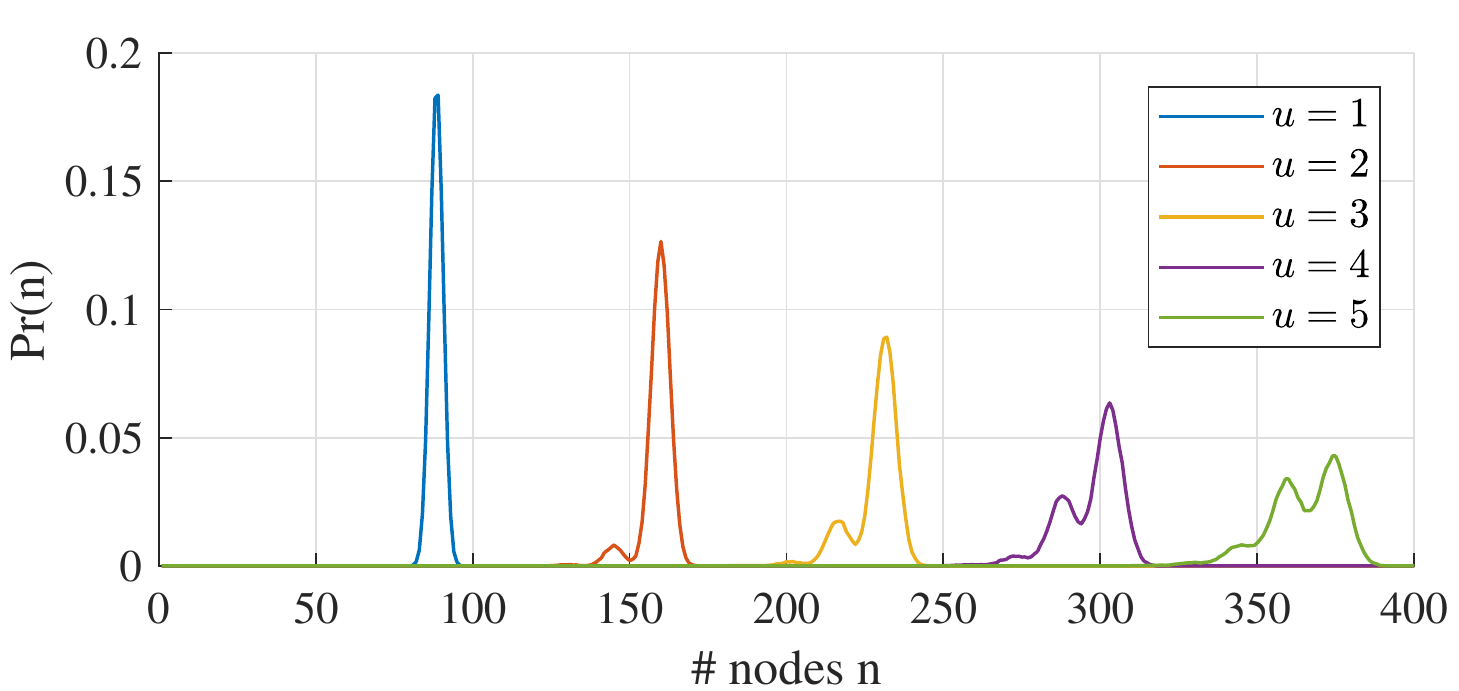}
  \caption{Probability distribution of the number of nodes in a PoMI, for different number of observed events $u$, obtained via experiment.}\label{fig:n_nodes}
\end{figure}

Since the analysis of the length of a Merkle-Patricia proof is based on the assumption that the tree is perfectly balanced, the analysis is validated by comparing analytical results both to numerical results obtained from a perfectly balanced tree, and to measurements obtained from the Merkle-Patricia tree implementation in PyEthereum, which is in general unbalanced, see Fig.~\ref{fig:ea}.
In particular, Fig.~\ref{fig:ea}(a) compares the analytical expression for the average number of nodes that compose a PoMI with the experimental data obtained from PyEthereum and with numerical results.
The results show that the analytical expression fits the numerical results obtained for a balanced tree; at the same time, it overestimates the average number of nodes, needed for a PoMI, in the Ethereum system.
This follows from the fact that the average depth of leaves in a Patricia tree is greater than in a balanced tree~\cite{rais1993limiting}, implying that some internal levels might not be completely populated, hence the slight reduction in number of nodes needed for the PoMI.
Fig.~\ref{fig:ea}(b) compares the length of the PoMI.
For the numerical and analytical results, each node is represented by the corresponding hash value, while for the experimental data the PoMI data structure is represented with Recursive Length Prefix (RLP)~\cite{wood2014ethereum}.
The RLP also contains information about the structure of the tree, therefore introducing a small overhead.
For this reason, the length for the experimental data is slightly larger than the ones of the numerical and analytical results.

The experimental setup also permits to characterize the distribution of the number of nodes in a PoMI, see Fig.~\ref{fig:n_nodes}, in which we show the results obtained for a blockchain with $\eta = 6$ levels, completely filled, therefore containing $L^6 = 16^6$ accounts, where $L$ is the maximum number of children of a node.
The relative position of the accounts in the tree clearly impacts the length of their PoMI and, hence, the communication cost of transmitting them.
In addition, the results provide insights on the consequence of using the expected length of PoMI, instead of its distribution, in \eqref{eq:p_a_distr}.
As the variance of the PoMI distribution remarkably increases with the number of included accounts, $u$, the precision of the approximation is decreased.
On the other hand, its contribution to the total length of the payload, $D_A$, is counterbalanced by the weight of accounts' data structure, which becomes predominant.
This is shown in details in the following text.

\subsection{Performance of the aggregation protocol}\label{sec:perf}

We consider a scenario in which the device is connected to BNs via a communication link parameterized as in Table~\ref{table:params}.
We remark that if the device is solely interested in observing accounts that are updated sporadically, the aggregation protocol only provides reduction of the communication overhead (the frame headers). Therefore, we focus the evaluation on the case where the device observes \emph{active} accounts. 
An account, $j$, is considered active if it is updated at least once every $T$ seconds, with probability $P_\text{A}$, i.e.
\begin{equation}\label{eq:active}
1 - \left( 1 - p_j \right)^{\left\lceil T/T_\text{B} \right\rceil} \geq P_\text{A}
\end{equation}
In the rest of the work, we set $P_\text{A} = 0.9$.
By requesting updates about more active accounts, than those of actual interest, the device can increase its privacy, at the cost of downloading unnecessary information. The application is further discussed in Sec.~\ref{sec:app}.

\begin{figure}[!tb]
\centering
\includegraphics[width=\columnwidth]{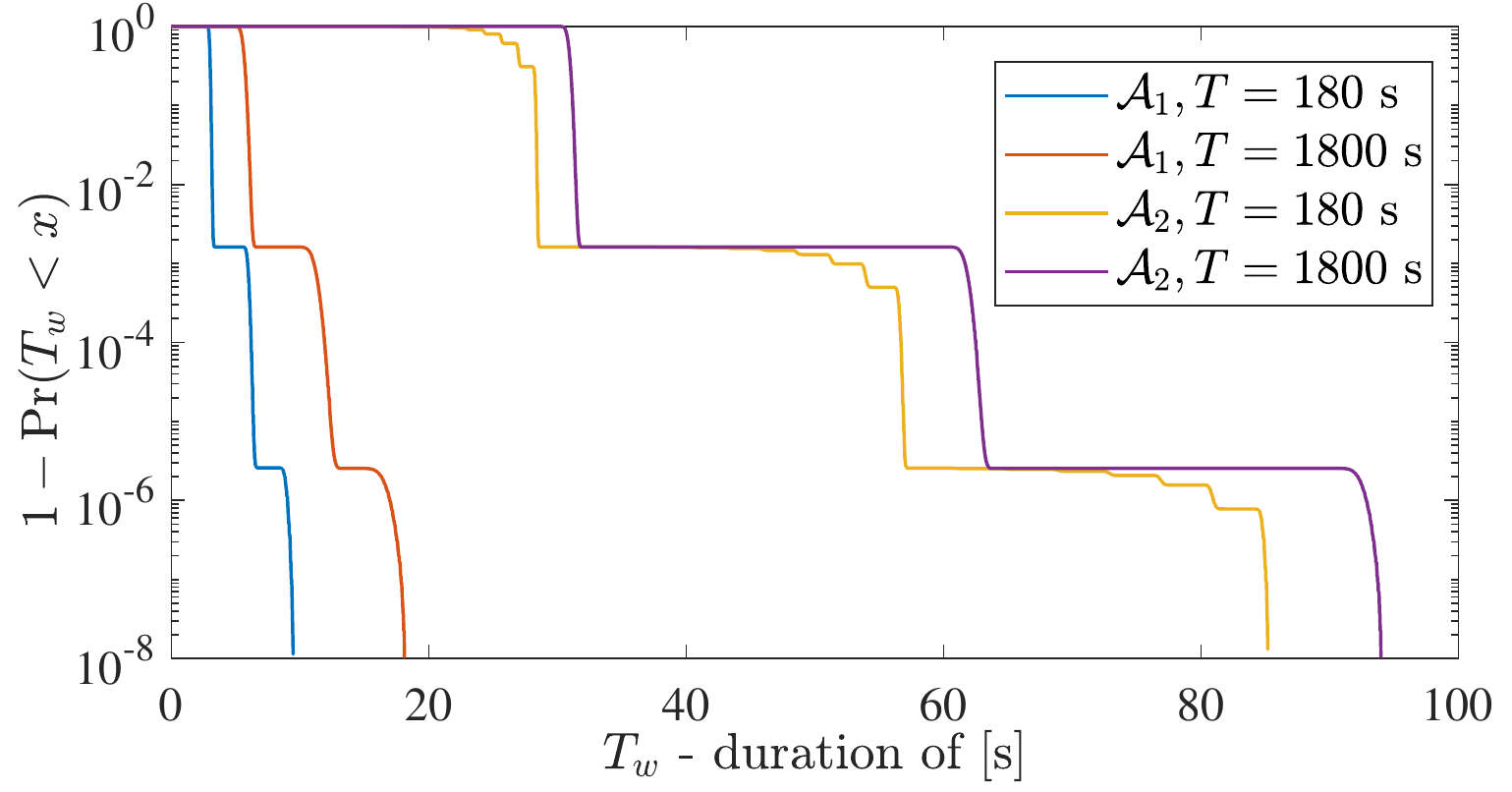}
  \caption{Complementary CDF of $T_w$ for two different intervals.}\label{fig:cdf}
\end{figure}
\begin{figure}[!tb]
\centering
  \subfloat[]{\includegraphics[width=\columnwidth]{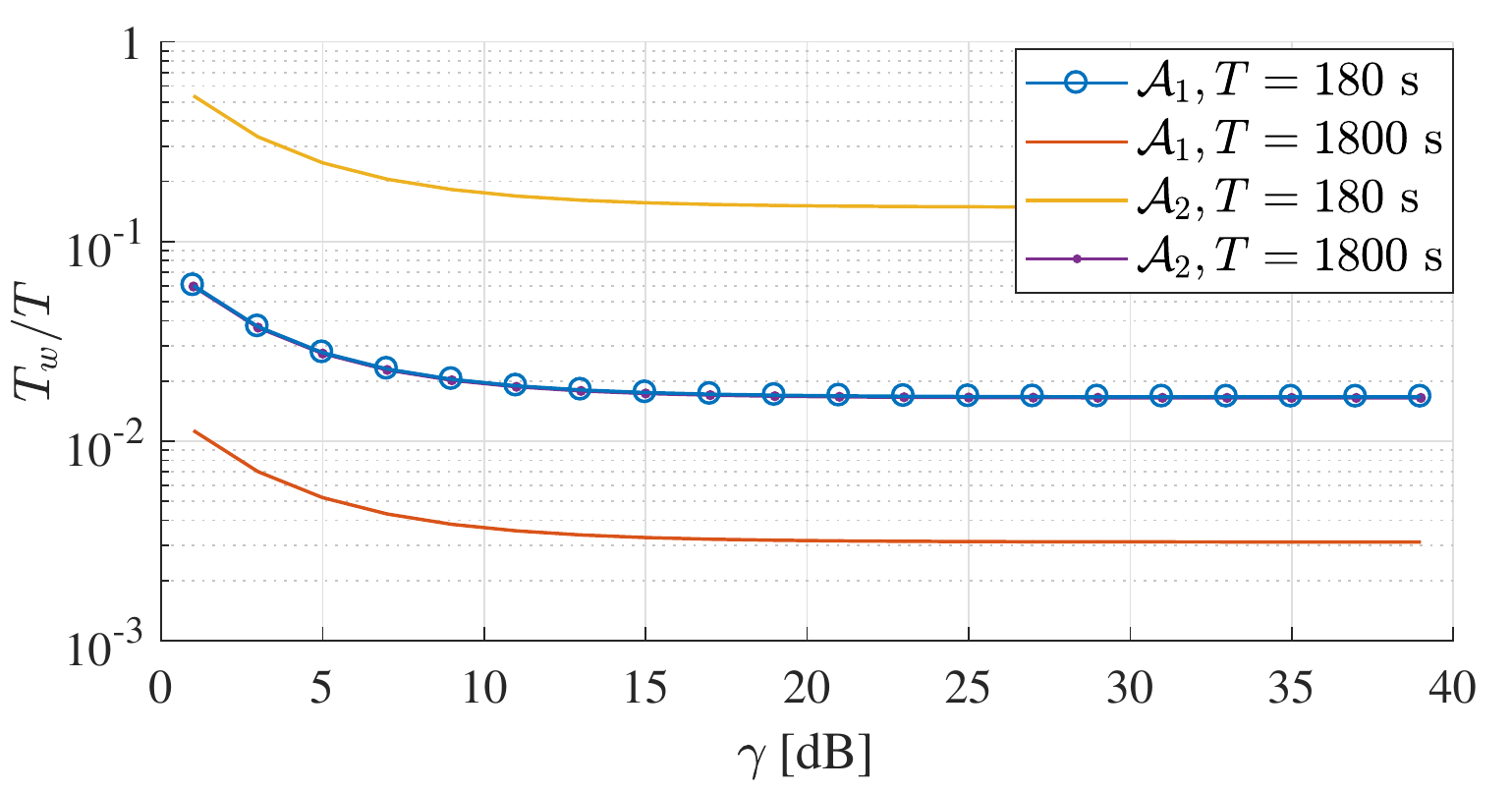}}
  
  \subfloat[]{\includegraphics[width=\columnwidth]{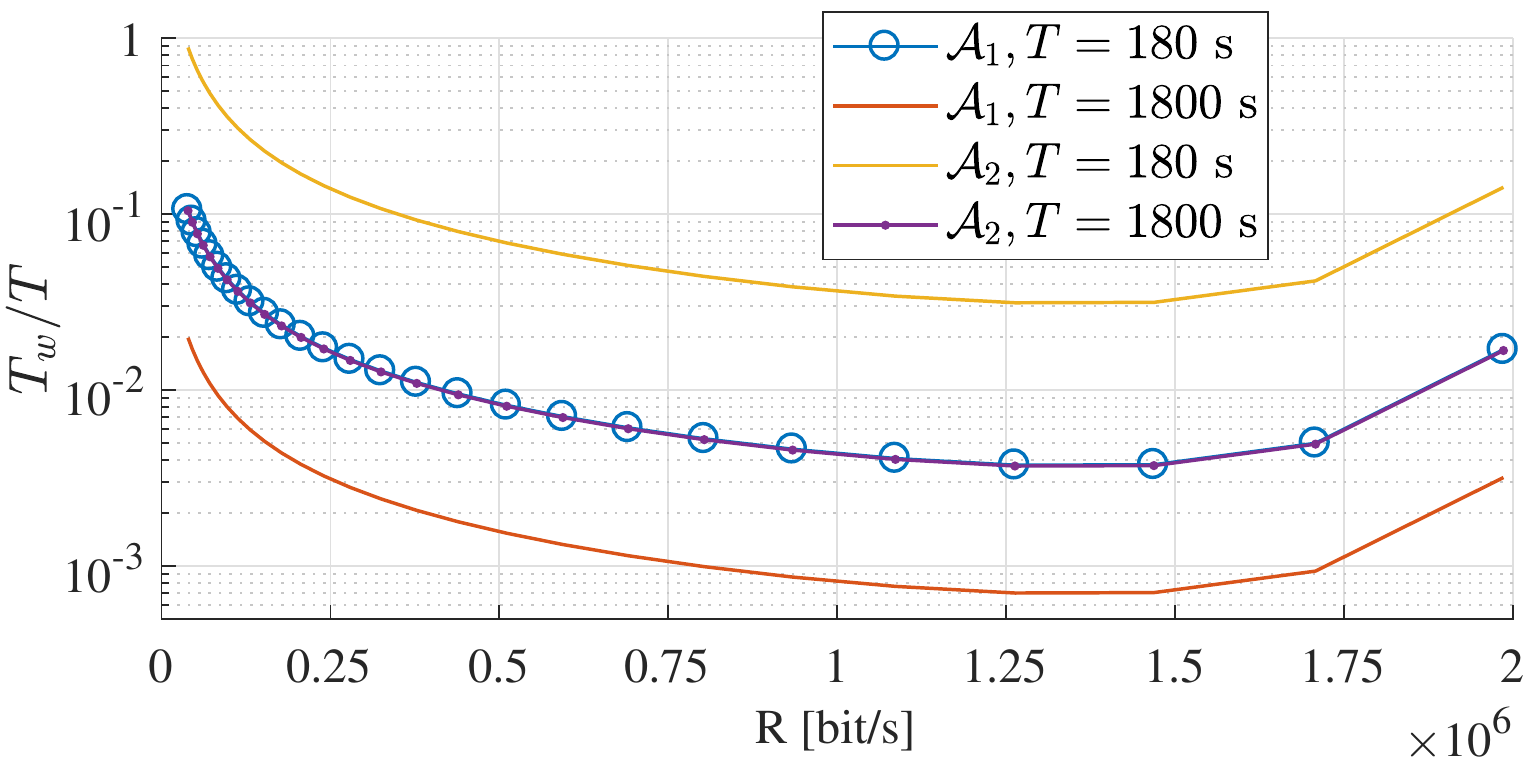}}
  \caption{Duty cycle of the device (a) for different values of SNR and (b) for different rates.}\label{fig:dc}
\end{figure}
\begin{figure}[!tb]
\centering
  \includegraphics[width=\columnwidth]{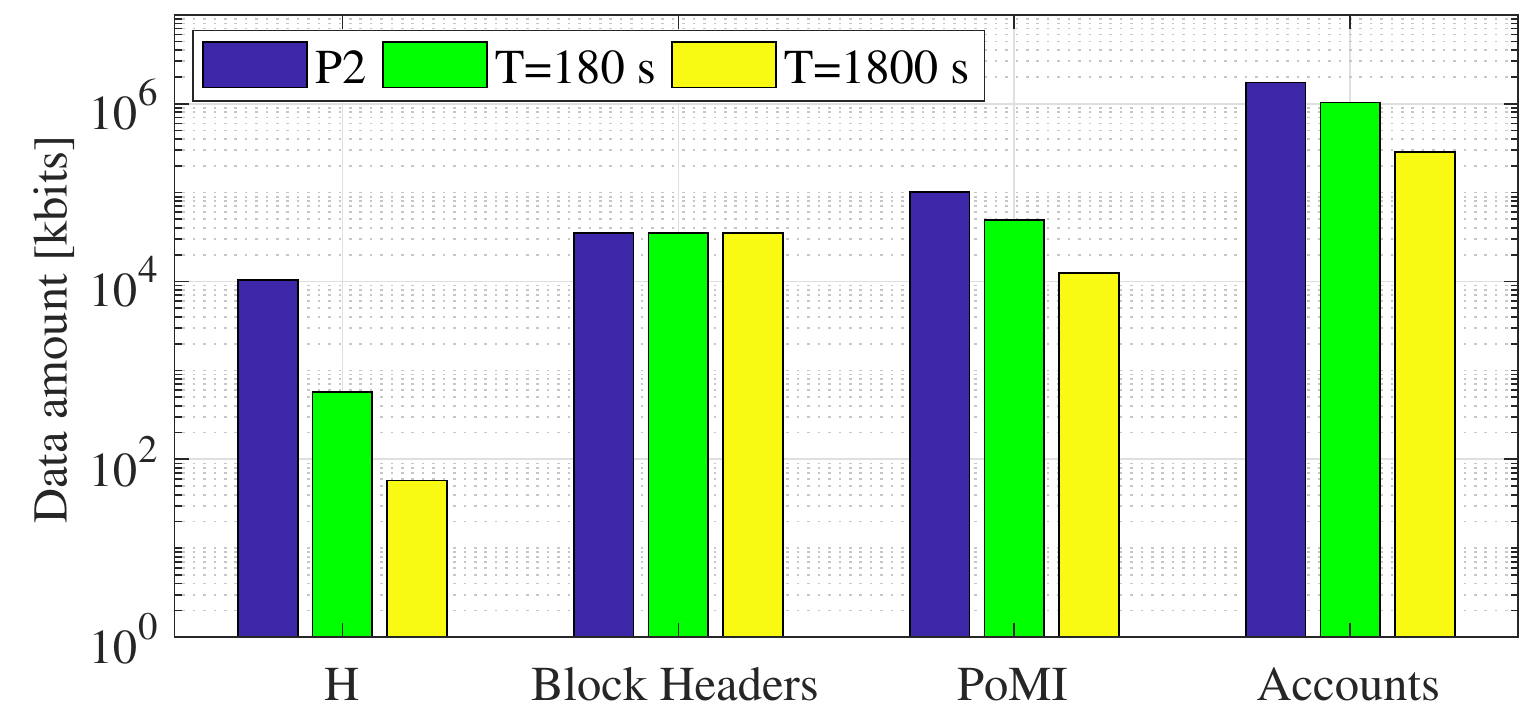}
    \caption{Amount of information downloaded for $\mathcal{A}_3$, during 24 hours, for protocol P2 and for protocol with aggregation with different values of $T$.}
    \label{fig:data}
\end{figure}
\begin{figure}[!tb]
\centering
\includegraphics[width=\columnwidth]{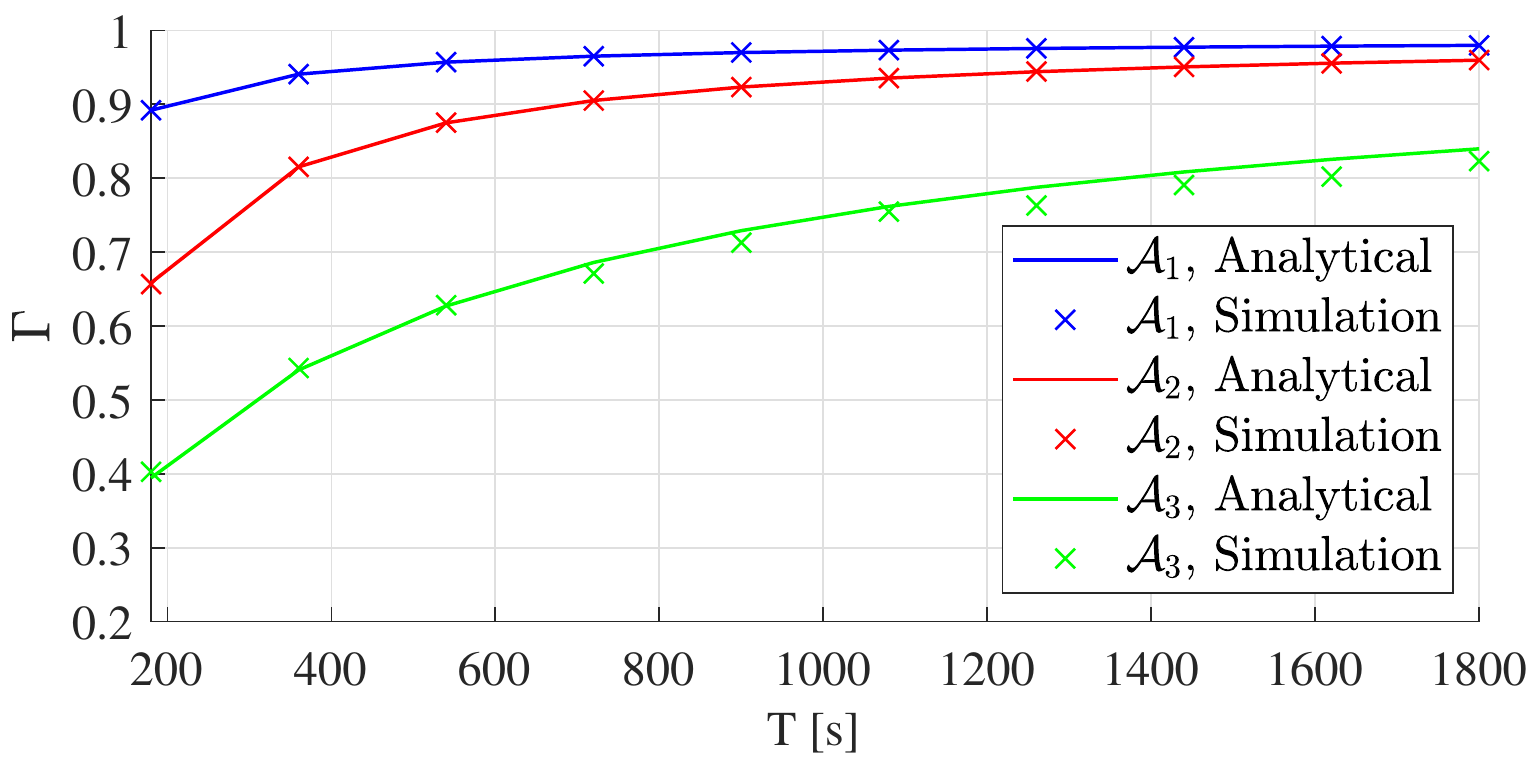}
  \caption{Gain of the aggregation protocol, analytical expression and simulation.}\label{fig:gamma}
\end{figure}

\subsubsection{Duty cycle trade-offs} Fig.~\ref{fig:cdf} reports the complementary CDF of the duration of the transmission, $T_w$, for two deterministic sets of observed accounts: $\mathcal{A}_1 = \{1, 2\}$, that contains the two most frequently updated ones, and $\mathcal{A}_2 = \{ j | 20 < j \leq 41\}$, containing the 20 less active accounts when $T=180$ s.\footnote{According to our definition, see \eqref{eq:active}, there are 41 active accounts for $T=180$ s and 695 for $T=1800$ s.} The sets are formed to illustrate two interesting limit scenarios.
The probability of channel outage, derived from $R$ and $\gamma$ of Table~\ref{table:params}, is $1.6 \cdot 10^{-3}$. The number of observed accounts, and their statistics, clearly plays a central role in shaping the CDF, as the size of the account data structure is much bigger than the size of a block header, see Table~\ref{table:params}.

The study of the duty cycle of the device, reported in Fig.~\ref{fig:dc}, covers several fundamental trade-offs.
Fig.~\ref{fig:dc}(a) shows that the duty cycle decreases when the channel quality (SNR) increases, due to the reduced number of retransmissions, and saturates for high values of SNR, as retransmissions are not likely to happen.
A possible strategy that can be adopted by the IoT device, to reduce its duty cycle, is to update less frequently, i.e. increase $T$, or reduce the set of observed accounts.
Fig.~\ref{fig:dc}(b) reports the duty cycle as function of the transmission rate of the wireless link. At low rates, the duty cycle of the device is drastically increased. On the other hand, selecting a high rate causes transmission failures and therefore retransmissions which become dominant when the rate reaches a certain level.

\subsubsection{Communication cost} The rest of the results focuses on how the different parts of the frame contribute to its total length and on the aggregation gain, defined in Sec.~\ref{sec:savings}.
We construct a set $\mathcal{A}_3$ containing $|\mathcal{A}_3| = 20$ accounts, by randomly selecting among those that are active during $T=1800$ s. It should be noted that $\mathcal{A}_2$ is a possible realization of $\mathcal{A}_3$.
Fig.~\ref{fig:data} shows the amount of information, downloaded during 24 hours of execution, for different values of $T$ and different realizations of accounts in $\mathcal{A}_3$.
In this scenario, there are no retransmissions; their effect would be a proportional increase in all the quantities.
The figure shows that most of communication cost is due to the size of the account data structures, which is an order of magnitude higher w.r.t. the size of PoMI, and two order of magnitudes higher than the size of the block headers and communication protocol headers.

The aggregation gain, $\Gamma$, is shown in Fig.~\ref{fig:gamma} for several values of the aggregation period, $T$, and compared with a simulation of the system. The figure shows a good match between the simulation and the analytics, and that the gain is remarkable, even for small values of $T$.
As $T\to\infty$ the observed accounts are updated almost surely during a period, and will be downloaded in the next transmission. This causes the gain to increase linearly when $T$ is large.

\subsection{Remarks on possibilities to further reduce of the communication cost}\label{sec:remark}

We briefly discuss possible directions for a further reduction of the communication cost for IoT lightweight clients, based on the insights provided by the evaluation of the protocol.
The size of the accounts' data structures has shown a prominent impact on the amount of transmitted data.
This can be reduced with several approaches, for example (i) by keeping their size as small as possible, during the design phase of the contract; (ii) by compressing the account information before sending it; (iii) by only sending the portion of account structure that has changed.

A completely different approach is to send the updates by means of the transactions tree, when the corresponding accounts are rarely updated. In fact, the size of a transaction is typically lower than the one of the account. In addition, while the state tree grows with the number of accounts, the transactions tree size is limited by the block size. 
This option, mentioned in Sec.~\ref{sec:syncprotocols}, has not been considered in this paper, as it does not provide aggregation gain.
Future works can consider this extension by (i) including in the system model the statistics of the number of transactions, that modify an account, in a single block; (ii) considering also the contribution of the transactions tree to the size of $P_{D_\text{A}}$ in \eqref{eq:ppab}; (iii) finding a strategy to decide if sending the update under the form of updated state, or as collection of transactions.

\subsection{Example application: privacy of IoT device}\label{sec:app}

We conclude this section by providing an example application of the aggregation protocol.
Consider a scenario in which the IoT device is solely interested in observing one (active) account, indexed as $j^{\star}$. However, to keep $j^{\star}$ private, it requests updates about additional accounts, in which it is not interested, from BNs (privacy by obfuscation) \cite{gruberunifying}. A malicious BN is aware that the IoT device is interested in one account and applies an outlier detection technique to find it. In the presented model, the only feature available to the BN is the relative frequency of update of accounts.
For both sides (device and BN), it is reasonable to assume that the set of observed accounts, indicated as $\mathcal{A}_4$, only contains active accounts, since non active accounts would be excluded by the outlier detection.

\begin{figure}[!tb]
\centering
\includegraphics[width=\columnwidth]{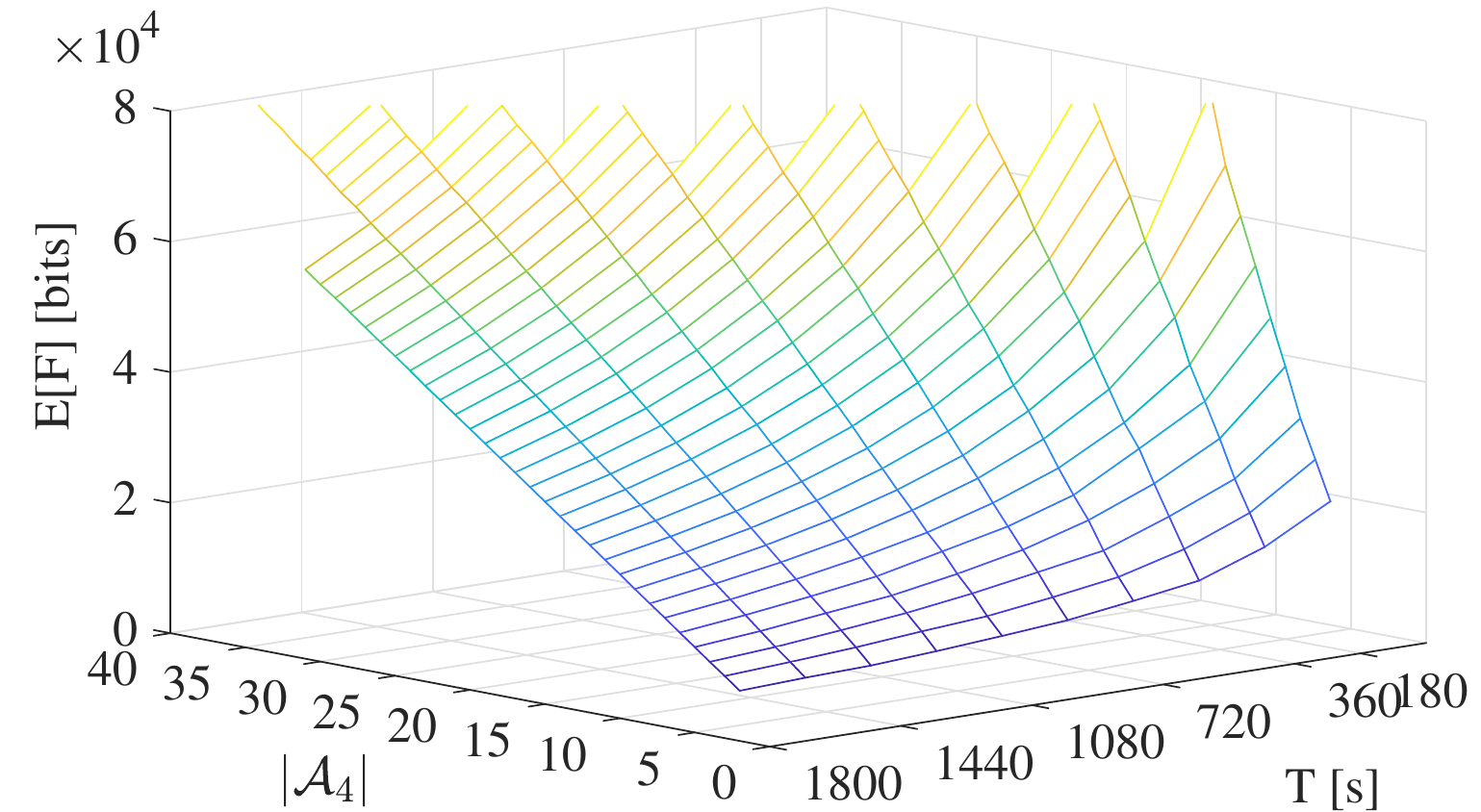}
  \caption{Communication cost as function of $|\mathcal{A}_4|$ and $T$.}\label{fig:privacy}
\end{figure}

Based on these considerations, the IoT device constructs $\mathcal{A}_4$ by adding $j^{\star}$ and other random active accounts. 
The construction starts with $\mathcal{A}_4 = \{ j^{\star} \}$, then $|\mathcal{A}_4|$ is iteratively incremented. At each iteration, the set of active accounts is split in $|\mathcal{A}_4|$ segments and one account is randomly picked from each segment ($j^{\star}$ is always picked among the accounts in its segment).
The iteration is repeated until the tolerated communication cost, expressed by \eqref{eq:savings}, is reached. Finally, $\mathcal{A}_4$ is sent to the BN.
There is a trade-off between the delay, given by the aggregation protocol, and privacy, i.e. $|\mathcal{A}_4|$. This trade-off is shown in Fig.~\ref{fig:privacy}, for different tolerated communication costs $\mathbb{E}[F]$, and $j^{\star} = 41$ (that is an active account).
A further improvement, not implemented in this paper, is impose that accounts in $\mathcal{A}_4$ should be located in proximity of each other in the state tree. In fact, this provides shorter PoMI and therefore lower communication cost, see Fig.~\ref{fig:n_nodes}.

\section{Conclusion}\label{sec:conclusion}

In this paper, we have investigated what is the communication cost of sending blockchain information to Ethereum-like lightweight clients.
A novel aggregation scheme has been proposed that has the potential to obtain a lower communication cost, at the expense of higher information delay, or availability of information, at the application layer.
The analysis of the scheme showed the probability distributions of the data structures exchanged over the wireless link, and their impact on the total downlink budget.

Finally, the results show that, if the statistics of account updates and the channel state are known, the lightweight clients can construct a list of events of interest that provides a predictable average communication cost. The example application illustrated how to apply our findings to improve the privacy of IoT devices.
The guidelines presented in this paper can be applied to design more advanced blockchain lightweight protocols.

\appendices

\section{Derivation of the expected number of nodes in a PoMI}\label{sec:appendix}

Under the relaxation described in Sec.~\ref{sec:analysis}, the probability that an arbitrary node at level $h$ is ancestor to one of $u$ modified leaf nodes (denoted by the binary random variable $X_h$) is $\Pr(X_h=1|N_{h-1}=n_{h-1},U=u)=(1-1/(Ln_{h-1}))^u$, where $N_{h-1}$ is the number of nodes at level $h-1$ that are ancestors to a modified leaf node and $L$ is the branching factor of the tree.
Since $X_h$ is a binary random variable, $\mathbb{E}_{X_h}[X_h|N_{h-1}=n_{h-1},U=u]=\Pr(X_h=1|N_{h-1}=n_{h-1},U=u)$, and the expected total number of ancestor nodes at level $h$ is $\mathbb{E}_{N_h}[N_h|N_{h-1}=n_{h-1},U=u]=Ln_{h-1}\cdot \mathbb{E}_{X_h}[X_h|N_{h-1}=n_{h-1},U=u]$. By the law of total expectation,
\begin{align*}
  \mathbb{E}_{N_h}\left[N_h|U=u\right]
  &=\mathbb{E}_{N_{h-1}}\left[\mathbb{E}_{N_h}\left[N_h|N_{h-1},U=u\right]|U=u\right]\\
  &=\mathbb{E}_{N_{h-1}}\left[LN_{h-1}\cdot \mathbb{E}_{X_h}\left[X_h|N_{h-1},U=u\right]|U=u\right]\\
  &=\mathbb{E}_{N_{h-1}}\left[LN_{h-1}\cdot \left(1-\frac{1}{LN_{h-1}}\right)^u\biggm| U=u\right].
\end{align*}
By applying a first-order Taylor expansion at $\mathbb{E}_{N_{h-1}}[N_{h-1}|U=u]$ we obtain
\begin{align}\label{eq:pomisinglelevel}
  \mathbb{E}_{N_h} & \left[N_h|U  =u\right]  \approx\\
 \nonumber  & L \, \mathbb{E}_{N_{h-1}} \left[n_{h-1}|U = u\right]\left(1-\frac{1}{L\mathbb{E}_{N_{h-1}}\left[n_{h-1}|U=u\right]}\right)^u.
\end{align}
Denoting $\bar{N}_{h}(u)=\mathbb{E}_{N_h}\left[N_h|U=u\right]$ and using the fact that the PoMI will always contain a single root to define $\bar{N}_{0}(u)=1$, \eqref{eq:pomisinglelevel} can be obtained by recursion.
The approximated number of nodes for a complete PoMI, in a tree of height $\eta$, is the sum of the nodes required at each level which yields \eqref{eq:n_eta}:
\begin{equation}\nonumber
\bar{N}_{\eta}(u) = \sum_{h=1}^{\eta} \bar{N}_h(u).
\end{equation}

\IEEEpeerreviewmaketitle

\nocite{*}
\bibliographystyle{IEEEtran}
\bibliography{references_rev}

\end{document}